\documentclass[12pt,a4paper, UKenglish]{article}
\usepackage[UKenglish]{babel}
\usepackage[UKenglish]{isodate}
\usepackage{sectsty}
\usepackage{xcolor}
\allsectionsfont{\color{blue}}

\usepackage{adjustbox}
\usepackage{natbib}
\usepackage[margin = 1in]{geometry}
\usepackage[small]{caption}
\usepackage{authblk}
\usepackage{multirow}
\usepackage{graphicx}
\usepackage{float} 
\usepackage[T1]{fontenc}
\usepackage{lmodern}
\usepackage{lmodern}
\usepackage{parskip}
\usepackage{caption}
\usepackage{subcaption}
\usepackage{amsmath}
\usepackage{accents}
\usepackage{multicol} 
\usepackage{comment}
\usepackage{setspace}
\usepackage[pdftex, bookmarks = true,colorlinks = true, allcolors = blue]{hyperref}
\cleanlookdateon
\usepackage[section]{placeins}

\cleanlookdateon

\begin{document}
	\title{\vspace{-2.0cm}\color{blue}On the unimportance of commitment for monetary policy}

\author{
\begin{tabular}{c}
Juan Paez-Farrell$^{*}$  \\
\end{tabular}}
\maketitle

\medskip
\begin{abstract}
\noindent In a New Keynesian model where the trade-off between stabilising the aggregate inflation rate and the output gap arises from sectoral asymmetries, the gains from commitment are either zero or negligible. Thus, to the extent that economic fluctuations are caused by sectoral shocks, policies designed to overcome the stabilisation bias are aiming to correct an unimportant problem.
\end{abstract}

\medskip
\vspace{0.2cm}
\vspace{10 cm}
\par\noindent\rule{0.5\textwidth}{0.5pt}\linebreak
\noindent \footnotesize $^*$ University of Sheffield, UK. E-mail \textcolor{blue}{j.paez-farrell@sheffield.ac.uk}. I am grateful for comments from Johannes Pfeifer and Christoph Thoenissen.

\clearpage
\newpage
\section{Introduction}
The standard New Keynesian (NK) model relies on understanding the dynamics of inflation and GDP on a single measure of aggregate output with real disturbances affecting supply and demand conditions for all individual goods in an equal manner. One of the key implications of the workshorse NK model is the 'divine-coincidence' \citep{BlanchardGali2007}, where stabilising the inflation rate automatically implies stabilisation of the output gap. Yet, a generally held view is that both variables cannot be simultaneously stabilised so that a trade-off exists. 

From a theoretical point of view, most of the contributions explaining this trade-off have centred on introducing a disturbance term to the New Keynesian Phillips curve (NKPC), such as including it exogenously \citep{CGG99} or by allowing for time-varying markups \citep{BenWoodford05}. Once the divine coincidence no longer holds, whether policy is time consistent has important welfare consequences \citep{SvenssonAER1997}. Even when the central bank does not wish to push output above potential, there is a stabilisation bias in that the inability to commit leads to an output gap that is more stable and an inflation rate that is more volatile, than is optimal.

The presence of the stabilisation bias in forward-looking rational expectations models has generated a large amount of research both on solving and on quantifying it. Potential solutions have included developing a reputation for commitment \citep{LevinePearlmanMcAdam}, speed-limit policies \cite{Walshspeed} or appointing a central banker with a high discount factor \citep{PaezFarrell12}, among others. Most attempts at quantifying the stabilisation bias have employed ad hoc loss functions and/or exogenous shocks to the NKPC such as \cite{DennisSoderstrom} and \cite{SchaumburgTambalotti}. However, when using ad hoc objectives the conclusions are very sensitive to the values assigned to the policy maker's preference parameters. Nonetheless, a general result that emerges is that the more forward-looking the model, the larger the benefits of implementing policy under commitment \citep{DennisSoderstrom}. Moreover, when maximising household welfare is used as the policy objective, the stabilisation bias can be large; using the model of \cite{SW2003}, \cite{LevinePearlmanMcAdam} find inflation equivalent gains of approximately $0.6\%$.

\begin{figure}[ht]	
    \centering
	\caption{Sectoral inflation rates}
	\hspace*{-1cm}	
\includegraphics[width=0.85\linewidth]{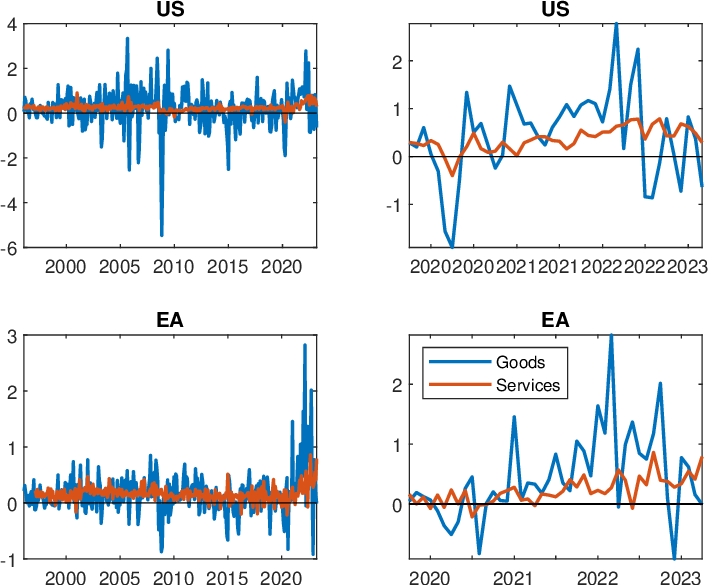}
\begin{minipage}{0.8\textwidth}
\bigskip
{\setstretch{1.0}\scriptsize{Note: The figures plot the monthly inflation rates in non-durable goods and services sectors for the US (top) and the euro area (EA, bottom). The plots on the right restrict the sample to the post-2018 period. US and euro area data were obtained from the \href{ https://fred.stlouisfed.org}{FRED} and from the \href{https://sdw.ecb.europa.eu/quickview.do?SERIES_KEY=122.ICP.M.U2.Y.GOODS0.3.INX&periodSortOrder=ASC}{ECB}, respectively.}}
\end{minipage}
\label{fig:Inflation}
\end{figure}
In this paper, we consider the gains from commitment in a standard New Keynesian model that includes two sectors. There are three key benefits to this approach. First, the 'shock' to the Phillips curve arises endogenously as a result of sectoral disturbances without relying on, for example, mark-up shocks.\footnote{While common practice as a way of introducing shocks in the Phillips curve, price and wage mark-up shock are often viewed as not truly structural. See \cite{Miguel} and \cite{ChariKehoeMcGrattan}.} Second, the use of two sectors to analyse monetary policy is particularly important in an open economy setting, where interest often lies in studying the consequences of shocks affecting the terms of trade. Lastly, the dynamics of inflation following the Covid-19 pandemic have led to a surge in interest on the effects of sectoral shocks as key drivers of inflation dynamics (\cite{Werningetal}, \cite{Fornaro}). As an example, Figure \ref{fig:Inflation} plots the evolution of monthly inflation in the non-durable goods and services sector over the period $1996M2-2023M4$. While the post-Covid-19 sample shows a dramatic difference in sectoral inflation rates, the whole sample suggests that focusing on economic activity at the sectoral level can yield important insights, even in relatively closed economies.

We find that in a standard two-sector New Keynesian model where the shock to the NKPC is caused by sectoral disturbances, the stabilisation bias is either zero or negligibly small. This surprising result is an implication of three key features of the standard two-sector NK model that are well known. First, full stabilisation of the sectoral inflation rates, the output gap and relative prices -- which would maximise social welfare -- is not possible unless prices are fully flexible in one sector \citep{Aoki2001}. If prices are equally sticky, the elasticity of sectoral inflation to the output gap is the same across sectors and we obtain an aggregate NKPC, so that stabilising the aggregate inflation rate is optimal as this also achieves output gap stabilisation. However, unlike in the one sector model, this does not imply full stabilisation as the relative price between the two sectors' goods will still be affected by the shock. Second, as the degree of price rigidity in one sector declines, so does its weight in the policy maker's objective function. Put differently, the policy maker should attach a greater weight to stabilising inflation in the sector with more sticky prices \citep{Benigno2004}. Lastly, a policy of stabilising a properly weighted index of the two sectors' inflation rates produces outcomes that are (almost) optimal. The key result of this paper is a simple implication of the findings above.

\section{A simple New Keynesian two-sector model}
We consider the two-sector New Keynesian model analysed by \citet[Chapter 3]{Woodfordb03}, which is a simple extension of the canonical NK model to two sectors, indexed by $j\in\{1,\, 2\}$. The consumption index $C_{t}$ that enters the utility function is a CES aggregate of $C_{1t}$, $C_{2t}$, where the elasticity of substitution between these two subindices is given by $\eta$. In addition, the weights attached to each subindex are subjected to positive shocks $\varphi_{1t}$ and $\varphi_{2t}$, respectively. This feature introduces sectoral asymmetries in the form of shifts in the  relative demands for the two sectors' outputs. Consequently, the demand for sectoral output will depend not only on the relative shock but also on the level of total expenditure $C_{t}$ as well as the sectoral relative price, $P_{jt}/P_{t}$, where $P_{t}$ is the overall price index.

Households supply labour of type $i$ to firms in sector $j$ and firms produce output via a function
\[y_{t}(i) = A_{jt}f(h_{t}(i))\]

Although the function $f(\cdot)$ is common across all firms, the presence of the sectoral technology shock $A_{jt}$ represents the second source of asymmetric disturnbances in the model.

Assuming Calvo pricing in each of the subsectors with parameter $\alpha_{j}$ gives rise to sectoral New Keynesian Phillips curves (NKPC) relating sectoral inflation, $\pi_{jt} \equiv \log\left(P_{jt}/P_{t,t-1}\right)$, to the aggregate output gap, $x_{t}$

\begin{equation}
\pi_{jt} = \beta E_{t}\pi_{j,t+1}+\kappa_{j} x_{t}+\gamma_{j}\left(\hat p_{Rt}-\hat p^{n}_{Rt}\right)
\label{eq:nkpcj}
\end{equation}
 
where $\beta$ denotes the household discount factor and the relative price gap is given by $\tilde p_{Rt}\equiv \left(\hat p_{Rt}-\hat p^{n}_{Rt}\right)$, with $\hat p_{Rt}\equiv \log(P_{2t}/P_{qt})$. The value of the latter under flexible prices is represented by $\hat p^{n}_{Rt}$. Equation (\ref{eq:nkpcj}) shows how the presence of asymmetric disturbances and price stickiness gives rise to the relative price gap, providing a structural rationale for the presence of a trade-off between stabilising the aggregate output gap and the aggregate inflation rate, which is given by

\begin{equation}
\pi_{t} \equiv n_{1}\pi_{1t}++n_{2}\pi_{2t}
\label{eq:pi}
\end{equation}

Here, $n_{j}$ represents the number of goods in sector $j$, so that $n_{1} +n_{2} = 1$. The elasticity $\pi_{jt}$ to the aggregate output gap, $\kappa_{j}$, varies across sectors only to the extent that the degrees of price rigidity, $\alpha_{j}$, differ. In the case, where $\alpha_{1} = \alpha_{2}$ the two NKPCs can be combined to produce an overall Phillips curve containing only the aggregate inflation rate and the aggregate output gap, thereby resulting in the divine coincidence.\footnote{Nonetheless, this does not imply that the sectoral output gaps, $x_{jt}$, or $\pi_{jt}$ and relative price gaps can all be stabilised.}

Given the definition of the relative price $\hat p_{Rt}$, its evolution is given by

\begin{equation}
\hat p_{Rt} = \hat p_{R,t-1}+\pi_{2t}-\pi_{1t}
\label{eq:pRt}
\end{equation}

Lastly, absent pricing frictions relative prices move in response to asymmetric shocks. That is
\begin{equation*}
\hat p^{n}_{Rt} = \frac{1}{\eta}\left[\left(\hat \varphi_{2t}-\hat \varphi_{1t}\right)-\left(\hat Y^{n}_{2t}-\hat Y^{n}_{1t}\right)\right]
\end{equation*}

where $Y^{n}_{jt}$ represents the flexible-price level of output in sector $j$. As in \cite{Woodfordb03}, we assume that the natural relative price follows an AR(1) process with persistence parameter $\rho$

\begin{equation}
\hat p^{n}_{Rt} = \rho \hat p^{n}_{R,t-1}+\varepsilon_{t}
\label{eq:pRn}
\end{equation}

From the process for $\hat p^{n}_{Rt}$ it can be seen that when the sectoral shocks are perfectly correlated, relative prices do not change so that the divine coincidence re-emerges. To complete the model, we represent the policy maker's objective as seeking to minimise the expected loss criterion

\begin{equation}
E[L] = (1-\beta)\sum_{t=0}^{\infty}\beta^{t}\left\{\sum_{j=1}^{2}w_{j}\pi_{jt}^{2}+\lambda_{x}x_{t}^{2}+\lambda_{R}\left(\hat p_{Rt}-\hat p^{n}_{Rt}\right)\right\}
\label{eq:loss}
\end{equation}

with weights $w_{j},\, \lambda_{x},\, \lambda_{R}>0$. This loss criterion can be regarded as a second order Taylor approximation to the lifetime utility of the representative household. Crucially, the weights on sectoral inflation are given by

\[w_{j} = n_{j}\frac{\kappa}{\kappa_{j}} \qquad \kappa \equiv \left[\frac{1}{\frac{n_{1}}{\kappa_{1}}+\frac{n_{2}}{\kappa_{2}}}\right]\]

Therefore, the more prices are flexible in a given sector, implying a lower $\alpha_{j}$ and larger $\kappa_{j}$ for an overall fixed level of $\kappa$, the lower the weight that the policy maker places on stabilising inflation in that sector.\footnote{See \citet[p.~437]{Woodfordb03}.} Moreover, the values above imply that $w_{1}+w_{2} = 1$.

\subsection{Parameterisation}
The parameter values chosen are based on \cite{RW98} as well as \cite{Woodfordb03} and are shown in Table \ref{t:calib}. Without loss of generality, we set $n_{1} = n_{2} = 1/2$ so that both sectors are of equal size, while the value of $\eta$ implies that the consumption aggregate is Cobb-Douglas. A steady state annualised interest rate of $4\%$ leads to the value shown for $\beta$.  Following \cite{Woodfordb03}, we consider the effects of variations in $w_{2}$, rather than in $\alpha_{j}$, for example. Doing so enables us to asses the effects of variations in the relative degree of price stickiness in each sector while maintaining the aggregate level (via $\kappa$) constant, noting that $w_{1} = 1-w_{2}$ so that the variations in $\alpha_{j}$ are implicit.\footnote{The values of $\lambda_{x}$ and $\lambda_{R}$, which depend on the structural parameters, are unaffected by the value of $w_{2}$.}

\begin{table}
\caption{Parameter values}
\begin{center}
\begin{tabular}{lllllllllllllllllll} 
\hline
$\beta$ & 0.99 & & $\eta$ &  1 & & $\kappa$ & 0.024 & & $n_{1}$ & 1/2 & $\lambda_{x}$ & 0.048 &&  $\lambda_{R}$ & 0.0288 && $\rho$ & 0.8 \\
\hline
\end{tabular}
\begin{minipage}{0.9\textwidth}
\scriptsize{Note: $\lambda_{x}$ and $\lambda_{R}$ depend on the other structural parameters of the model but are invariant to $\omega_{2}$.}
\end{minipage}
\end{center}
\label{t:calib}
\end{table}

\subsection{Results}\label{sec:Woodfordresults}
The model equations consist of (\ref{eq:nkpcj})-(\ref{eq:pRn}) plus a characterisation of monetary policy. For the latter we consider three alternatives: discretion, commitment and a policy of stabilising a weighted average of the sectoral inflation rates

\begin{equation}
\pi^{T} = \phi \pi_{1t}+(1-\phi)\pi_{2t} \qquad 0\leq \phi \leq 1
\label{eq:piT}
\end{equation}

\begin{figure}[ht]	
    \centering
		\caption{Gains from commitment}
	\hspace*{-1cm}	
     \begin{subfigure}[b]{0.4\textwidth}
         \centering
         \includegraphics[width=\textwidth]{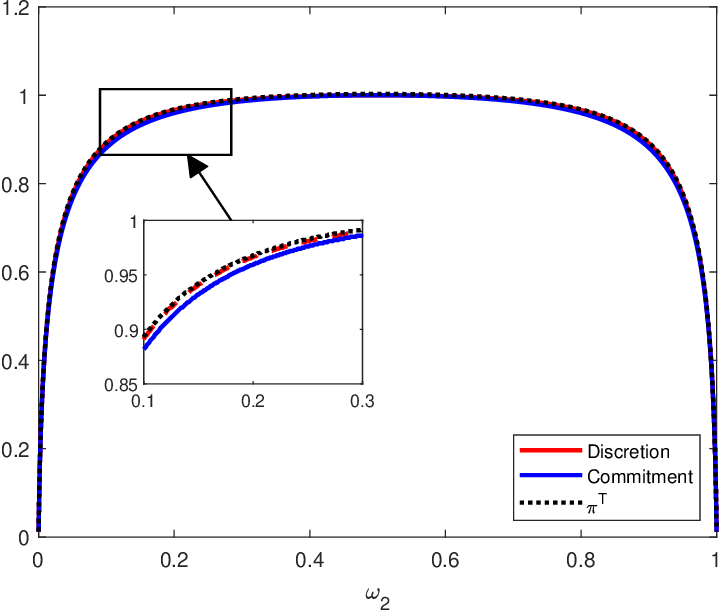}
         \caption{Losses for each policy.}
         \label{fig:loss_comparison}
     \end{subfigure}
     \begin{subfigure}[b]{0.4\textwidth}
         \centering
         \includegraphics[width=\textwidth]{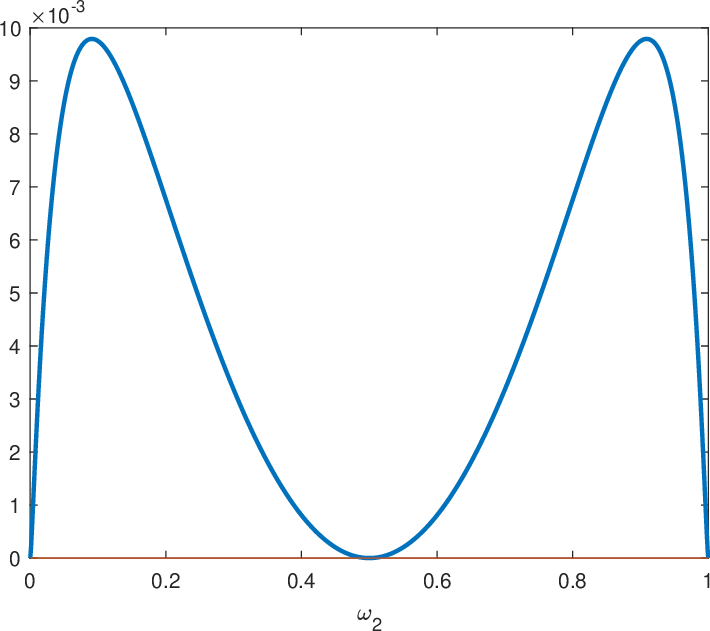}
         \caption{Stabilisation bias}
         \label{fig:stab_bias}
     \end{subfigure}
	\begin{minipage}{0.8\textwidth}
\bigskip
{\scriptsize Note: Subplot \ref{fig:loss_comparison} shows the losses normalised so that at $\omega_{2}=0.5$ those under commitment equal one, where $\omega_{2}$ captures the degree of price stickiness in sector $2$ relative to that in sector $1$. Subplot \ref{fig:stab_bias} shows the stabilisation bias (loss under discretion minus under commitment).}
\end{minipage}
\label{fig:1}
\end{figure}

In all cases the monetary authority policy minimises (\ref{eq:loss}) so that under the policy described by equation (\ref{eq:piT}) the parameter $\phi$ is chosen optimally. The results can be seen in Figure \ref{fig:1}, which show that the losses generated by all three policies are almost identical, accounting for the minute size of the stabilisation bias. 
From panel \ref{fig:loss_comparison}, several features are worth highlighting. First, for $\omega_{2} = \{0,\, 1\}$ losses are zero as in these cases prices are sticky in only one sector and therefore the divine coincidence arises. Second, for $\omega_{2} = 0.5$, both sectors are equally sticky and stabilising the aggregate inflation rate is optimal as this policy also stabilises the output gap. However, under such an equilibrium the relative price gap ($\tilde p_{Rt}$) is not stabilised which, given the parameterisation above, leads to the largest losses. Crucially, at these three points the welfare gains in moving from commitment to discretion are zero and therefore so is the stabilisation bias. The reason for this result is that at $\omega_{2} = \{0,\, 0.5,\, 1\}$ fully stabilising $\pi^{T}$ is optimal so that there is no trade-off involved. Put differently, whenever stabilising $\pi^{T}$ is optimal, discretion and commitment generate the same outcomes. Thirdly, for all other values of $\omega_{2}$, weighted inflation targeting is very close to optimal. 

A consequence of the above is that regardless of the differing degrees of relative price stickiness, captured by $\omega_{2}$, the gains in moving from commitment to discretion are trivial, which are shown in panel \ref{fig:stab_bias}. The intuition behind this result is very simple and hinges on the near-optimality of targeting $\pi^{T}$: if aiming to fully stabilise a single variable every period is quasi-optimal, then the policy maker does not face an intertemporal trade-off. The stabilisation bias arises due to the policy maker's inability to commit and the effect this has on private sector expectations. However, in the two-sector model considered above, a monetary authority operating under discretion will aim for a policy that will closely resemble setting $\pi^{T}$ to zero every period, so that there is no gain in affecting $E_{t}\pi_{j,t+1}$ following a shock. 

\begin{figure}[ht]	
    \centering
	\caption{Gains from commitment}
	\hspace*{-1cm}	
     \begin{subfigure}[b]{0.45\textwidth}
         \centering
         \includegraphics[width=\textwidth]{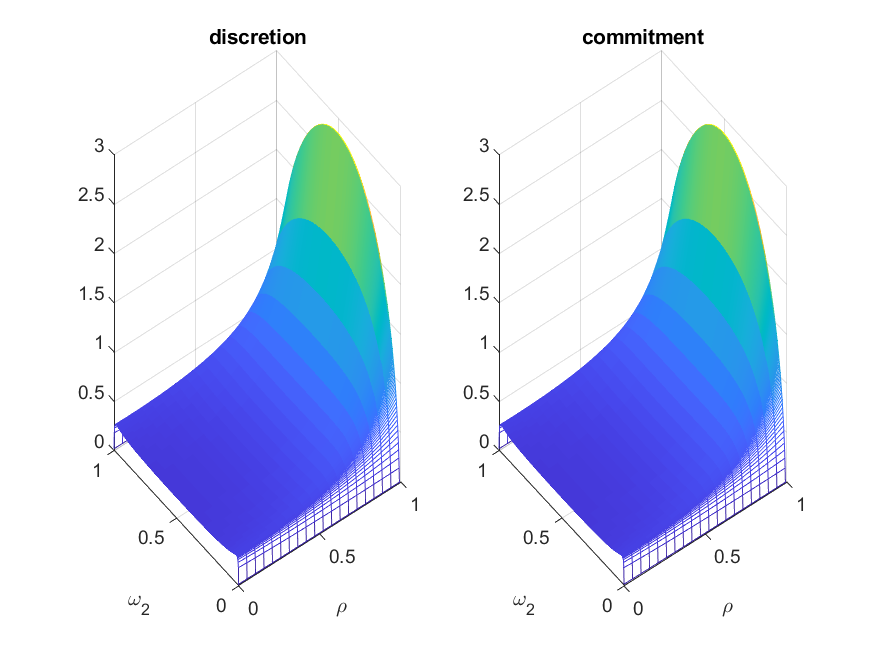}
         \caption{Losses for each policy.}
         \label{fig:loss_comparison2}
     \end{subfigure}
     \begin{subfigure}[b]{0.45\textwidth}
         \centering
         \includegraphics[width=\textwidth]{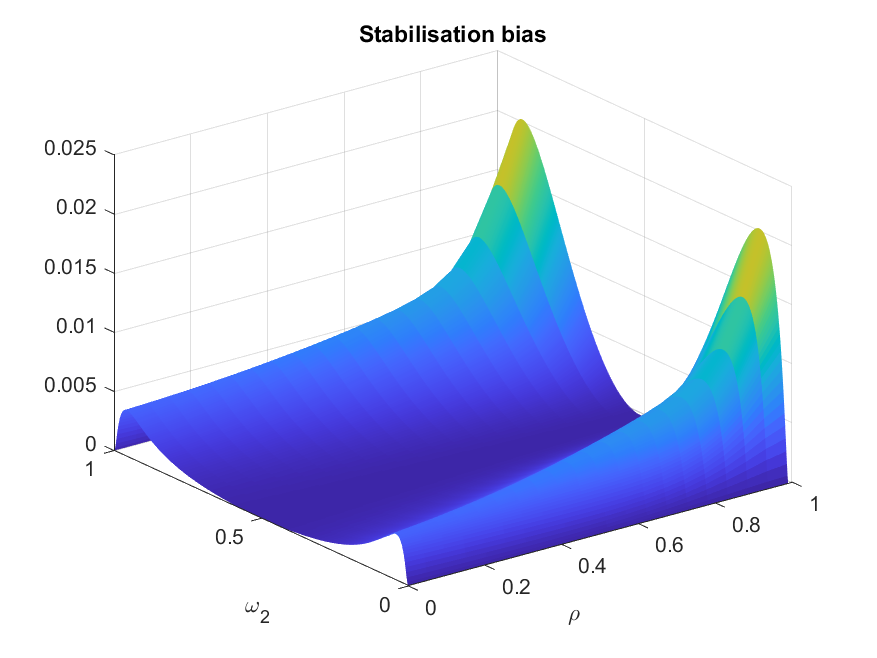}
         \caption{Stabilisation bias of discretion.}
         \label{fig:stab_bias2}
     \end{subfigure}
	\begin{minipage}{0.85\textwidth}
\bigskip
{\scriptsize{Note: Losses are given by the vertical axes. Subplot \ref{fig:loss_comparison2} shows the losses normalised so that at $\omega_{2}=0.5,\,\rho = 0.8$ those under commitment equal one. Subplot \ref{fig:stab_bias2} shows the stabilisation bias (loss under discretion relative to commitment).}}
\end{minipage}
\label{fig:2}
\end{figure}

Further insights can be gleaned by assessing to what extent the effects are sentitive to the persistence of the natural relative price deviations, $\rho$, which is shown in Figure \ref{fig:2}. Even for very large values of $\rho$ ($0.99$), the stabilisation bias remains negligible, which also implies that a monetary policy of targeting the weighted inflation index in (\ref{eq:piT}) performs remarkably well across the parameter space.

\begin{figure}[ht]	
    \centering
	\caption{Gains from commitment (inflation equivalent)}
	\hspace*{-1cm}	
\includegraphics[width=0.7\linewidth]{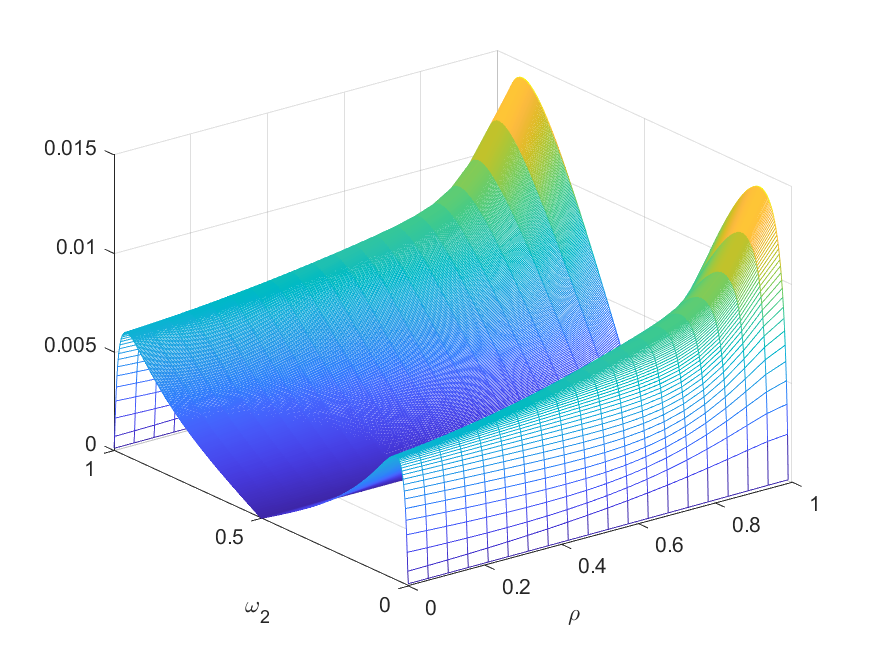}
\begin{minipage}{0.85\textwidth}
\bigskip
{\scriptsize{Note: The inflation equivalent, given by equation (\ref{eq:piequiv}), measures the inefficiency of discretionary monetary policy as the permanent deviation of inflation from its target that in welfare terms is equivalent to transitioning from discretion to commitment.}}
\end{minipage}
\label{fig:piequivalentl}
\end{figure}

One may wonder whether results above hinge on the metric used to measure the stabilisation bias given that the loss differential has no straightforward economic interpretation. Therefore, following \cite{Jensen2002} and \cite{DennisSoderstrom}, we calculate the permanent deviation of aggregate inflation that yields the same welfare gains as moving from discretion to commitment. For the model considered in this paper, this implies an increase in both $\pi_{1}$ and $\pi_{2}$ of the same magnitude, hence resulting in an identical rise in $\pi$, which is denoted by $\hat \pi$. This quantity is given by

\begin{equation}
\hat \pi  = \sqrt{L_{d}-L_{c}}
\label{eq:piequiv}
\end{equation}

where $L_{d}$ and $L_{c}$ denote the losses under discretion and commitment, respectively.\footnote{If both $\pi_{1}$ and $\pi_{2}$ increase by the same amount and with $\omega_{1}+\omega_{2} = 1$, the inflation equivalent is defined by $L_{c}+(1-\beta)\sum_{t=0}^{\infty}\beta^{t}\hat \pi^{2} = L_{d}$.} A further benefit of assessing the gains from commitment using  this measure is that it makes comparison with the related literature possible. Using the metric (\ref{eq:piequiv}) does not overturn the findings above; if anything, Figure \ref{fig:piequivalentl} shows that it reinforces the finding that that the inefficiency of discretionary policy is trivially small. For example, at the default value of $\rho = 0.8$ considered by \cite{Woodfordb03}, the absence of commitment entails a welfare loss equivalent to a permanent increase in inflation of $0.01$ points.
 
How do the results above compare to previous findings in the literature? A general conclusion that emerges is that the values obtained have been much larger than those reported in this paper. \cite{DennisSoderstrom} consider two estimated models with strong forward-looking features, \cite{FuhrerMoore} and \cite{OrphanidesWieland}, obtaining inflation equivalent values ranging from 0.3 to 0.6. \cite{Lees2007} and \cite{Liu2007} estimate DSGE models on New Zealand data and both find that the inflation equivalent gains from commitment, using ad hoc objectives, range from 0.7 to 4. A somewhat larger value of 4.6 is obtained by \cite{Tillmann} using a simple parameterised closed-economy one-sector New Keynesian model. However, in his case the comparison hinges between discretion and a policy of robust delegation when the degree of persistence in the shock to the Phillips curve is unknown.

Overall, the model considered in this paper suggests that when a trade-off between output gap and inflation stabilisation arises due to the presence of asymmetric disturbances and welfare is measured in a model-consistent manner, a discretionary policy maker delivers outcomes that are as good as those obtained under commitment. Alternatively, committing to a policy of fully stabilising a weighted inflation measure given by (\ref{eq:piT}) is also optimal.

\section{The case of the open economy}
As a robustness check and because the inclusion of sectoral asymmetries is often regarded as crucially important in analysing monetary policy in open economies, we extend the analysis carried out in the previous section to the model in \cite{LiuPappa2008}. This consists of a two-country New Keynesian model that includes tradable and non-tradable sectors. Utility is logarithmic in the final consumption good ($C_{t}$), which is a Cobb-Douglas (C-D) aggregate of traded ($T$) and non-traded ($N$) goods, with the share of the former denoted by $\alpha$. Furthermore, $C_{Tt}$ is in turn a C-D aggregate of the domestically-produced good ($H$) and the imported good ($F)$, where the share of the former, $\omega$, measures the degree of home bias.\footnote{These assumptions imply that under flexible prices labour is constant so that each sector's level of output move one-for-one with the technology shocks in that sector.} In this model, the source of asymmetric disturbances lies in the presence of sector-specific technology shocks in the tradable and non-tradable sectors in each country, denoted by $A_{Tt}$ and $A_{Nt}$, respectively. The key equations (in log-linear form) describing the model for the Home economy are then given by

\begin{align}
\pi_{Nt} &= \beta E_{t}\pi_{N,t+1}+\kappa_{N}x_{Nt}
\label{eq:LPpin}\\
\pi_{Ht} &= \beta E_{t}\pi_{Ht+1}+\kappa_{T}x_{Tt}
\label{eq:LPpih}\\
\Delta x_{Nt} &= \Delta x_{Tt}-\pi_{Nt}+\pi_{Ht}-\Delta \hat a_{Nt}+\Delta \hat a_{Tt}
\label{eq:LPexpenditure}
\end{align}

where $\hat a_{jt}$ is technology in sector $j$. The first two equations are the New Keynesian Phillips curves in the $N$ and $T$ sectors, respectively. Given the assumption of C-D aggregation, expenditure changes in the two sectors are proportional to each other and this captured by equation (\ref{eq:LPexpenditure}).

Assuming that monetary policy operates under a cooperating regime and that both countries are of equal size, social welfare is given by
\begin{equation}
\begin{split}
W = -\frac{1}{4}E_{0}\sum_{t=0}^{\infty}\beta^{t}&\biggl\{(1-\alpha)\left(x_{Nt}^{2}+\frac{\theta_{N}}{\kappa_{N}}\pi^{2}_{Nt}\right)+\tilde \alpha \left(x_{Tt}^{2}+\frac{\theta_{T}}{\kappa_{T}}\pi^{2}_{Ht}\right)\\
&+\left(1-\alpha^{\ast}\right)\left(x_{Nt}^{\ast2}+\frac{\theta^{\ast}}{\kappa_{N}^{\ast}}\pi_{Nt}^{\ast2}\right)+\tilde\alpha^{\ast}\left(x_{Tt}^{\ast2}+\frac{\theta^{\ast}}{\kappa_{T}^{\ast}}\pi_{Ft}^{\ast2}\right)\biggr\}
\end{split}
\label{eq:LiuPappaW}
\end{equation}

where asterisks denote foreign variables and with $\theta_{j},\, j=\{N,T\}$, representing the elasticity of substitution between the differentiated goods in sector $j$.\footnote{\cite{LiuPappa2008} also consider policy under a Nash regime. For the purposes of this paper, comparing commitment to discretion, it makes no difference what regime is implemented.} The weighting placed on each sector is proportional to the sector's size. This is $(1-\alpha)$ for the non-tradable sector while that for tradables is adjusted to take into account that the planner internalises the terms of trade externality so that. Moreover, the more that prices are flexible in a given sector the lower the weight that the planner places on stabilising inflation in that sector.

In parameterising the model, we follow the values used by \cite{LiuPappa2008}, which are shown in Table \ref{t:calibLP}. Although some parameters are not structural, such as $\kappa_{j}$, the analysis conducted below takes this into account.

\begin{table}
\caption{Parameter values}
\begin{center}
\begin{tabular}{llllllllllllllll} 
\hline
$\beta$ & 0.99 & & $\alpha$ &  0.3 & & $\omega$ & 0.7 & & $\theta$ & 10 & $\kappa_{j}$ & 0.0858 &&  $\tilde \alpha$ & 0.3  \\
\hline
\end{tabular}
\vskip.2\baselineskip
\begin{minipage}{0.75\textwidth}
\scriptsize{Note: The subscript $j$ refers to sector $N$ or $T$. Identical values are assumed for the foreign economy.}
\end{minipage}
\end{center}
\label{t:calibLP}
\end{table}

\begin{figure}[ht]	
    \centering
	\caption{Losses in the \cite{LiuPappa2008} model}
	\hspace*{-1cm}	
\includegraphics[width=0.7\linewidth]{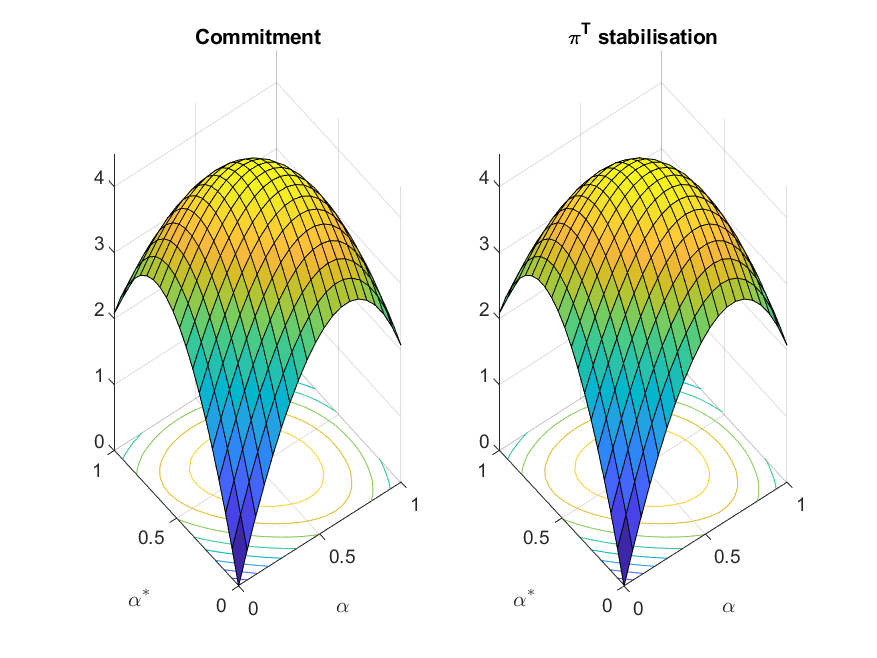}
	\begin{minipage}{0.8\textwidth}
\bigskip
{\scriptsize{Note: Losses are reported as a percentage of steady state consumption. The value of $\alpha$ ($\alpha^{\ast}$) represents the size of the tradable sector. Losses under discretion are not reported as they are identical to those under commitment.}}
\end{minipage}
\label{fig:liupappaalphascommosr}
\end{figure}

\begin{figure}[ht]	
    \centering
	\caption{Optimal $\phi$ in the \cite{LiuPappa2008} model}
	\hspace*{-1cm}	
\includegraphics[width=0.7\linewidth]{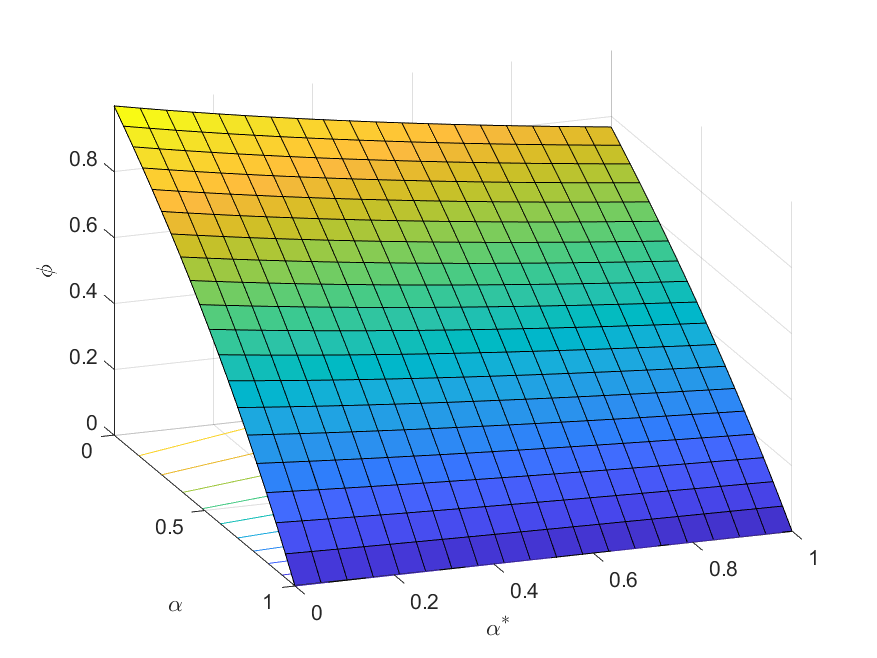}
\begin{minipage}{0.65\textwidth}
\scriptsize{Note: The figure plots the optimal value of $\phi$ for the case where the social planner stablises $\pi^{T}_{t}$ in equation (\ref{eq:piTLP}).}
\end{minipage}
\label{fig:liupappaphioptalphas}
\end{figure}

We again solve the model under discretion, commitment and the stabilisation of an inflation index, $\pi^{T}_{t}$. For this model, the latter is given by
\begin{equation}
\pi^{T}_{t} = \phi \pi_{Nt}+(1-\phi)\pi_{Ht}
\label{eq:piTLP}
\end{equation}

and an analogous index is used for the foreign economy. The losses as we alter the size of the tradable sector in either country are shown in Figure \ref{fig:liupappaalphascommosr} where for comparability with \cite{LiuPappa2008}, the losses are reported as a proportion of steady state consumption. The first thing to note is that losses under discretion are not reported as they are almost identical to those under commitment and the reason is explained by the right panel. It shows that when the policy maker stabilises an inflation index in each country described by (\ref{eq:piTLP}), the resulting outcomes are identical to those under commitment, so that the stabilisation bias is zero. Further detail is provided by Figure \ref{fig:liupappaphioptalphas}, which shows how the optimal value of $\phi$ changes with the values of $\alpha$ and $\alpha^{\ast}$.\footnote{As the figure for $\phi^{\ast}$ is analogous, it is not reported.} As the size of the tradable sector ($\alpha$) increases, the optimal value of $\phi$ falls monotonically, albeit non-linearly. These responses are consistent with those obtained from implementing (\ref{eq:piT}) in the model by \cite{Woodfordb03} considered earlier and by \cite{Benigno2004}.

We can additionally conduct the analysis in Section \ref{sec:Woodfordresults} of altering the degree of relative degrees of price rigidities whilst maintaining the overall aggregate measure intact. To do this, we can re-write the welfare objective (\ref{eq:LiuPappaW}) as

\begin{equation}
\begin{split}
W  = -\frac{1}{4}\theta E_{0}\sum_{t=0}^{\infty}\beta^{t}&\biggl\{\frac{1}{\kappa}\left(w_{1}\left(\frac{\kappa_{N}}{\theta}x^{2}_{Nt}+\pi^{2}_{Nt}\right)+w_{2}\left(\frac{\kappa_{T}}{\theta}x^{2}_{Tt}+\pi_{Ht}^{2}\right)\right)\\
&+\frac{1}{\kappa^{\ast}}\left(w_{1}^{\ast}\left(\frac{\kappa_{N}^{\ast}}{\theta}x^{\ast2}_{Nt}+\pi^{\ast2}_{Nt} \right)+w_{2}^{\ast}\left(\frac{\kappa_{T}^{\ast}}{\theta}x^{\ast2}_{Tt}+\pi^{\ast2}_{Ft} \right) \right)\biggr\}
\end{split}
\end{equation}

where 
\[w_{1} = (1-\alpha)\frac{\kappa}{\kappa_{N}} \qquad w_{2} = \tilde\alpha \frac{\kappa}{\kappa_{T}} \qquad \kappa \equiv  \left[\frac{(1-\alpha)}{\kappa_{N}}+\frac{\tilde\alpha}{\kappa_{T}}\right]^{-1}\]

implying that $w_{1}+w_{2} = 1$ (the equivalent expressions for the foreign economy are analogous). We can then determine the potential gains of moving from discretion to commitment as the degree of relative price flexibility varies across sectors. For example, for a given $\kappa$, as the prices in the tradable sector become more flexible, $\kappa_{T}\rightarrow \infty$, leading to $w_{1} = 1$, $w_{2} = 0$ and $\kappa_{N} = (1-\alpha)\kappa$. However, unlike the model considered previously, the fact that the two sectors are not of equal size ($\alpha<0.5$) coupled with the assumption of coordination, implies that when price stickiness is the same across both the $N$ and $T$ sectors in the home economy the values of the weigths in the welfare function are given by 

\[w_{1} = \frac{1-\alpha}{1-\alpha+\tilde \alpha} \qquad w_{2} = \frac{\tilde \alpha}{1-\alpha+\tilde \alpha}\]

\begin{figure}[ht]	
\begin{center}
	\caption{Stabilisation bias in the \cite{LiuPappa2008} model}
	\hspace*{-1cm}	
\includegraphics[width=0.7\linewidth]{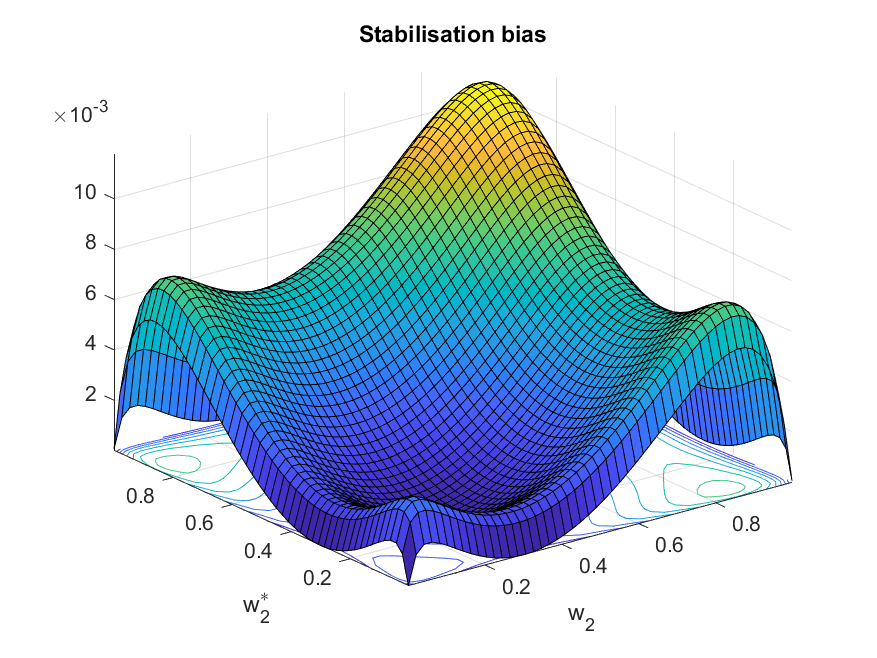}
\begin{minipage}{0.8\textwidth}
\scriptsize Note: The figure plots the losses under discretion minus those under commitment (as a percentage of steady state consumption) as the relative degrees of price stickiness in the Home and Foreign tradable sectors, $w_{2}$ and $w_{2}^{\ast}$, respectively, vary. When the tradable and non-tradable sector have the same degree of price flexibility the value of $w_{2} = 0.3$
\end{minipage}
\label{fig:liupappaw2sstabilbias}
\end{center}
\end{figure}

Using the parameterisation in Table \ref{t:calibLP} then implies $w_{1} = 0.7$ and $w_{2} = 0.3$. Figure \ref{fig:liupappaw2sstabilbias} plots the stabilisation bias across all the values of $w_{2}$ and $w_{2}^{\ast}$, once again highlighting how a move form discretion to commitment has a negligible impact on welfare. In this open economy model, fluctuations in the terms of trade act as a relative price shock that introduces a trade-off between stabilising the output gaps and inflation rates, even when the degrees of price stickiness are the same across sectors. Nonetheless, the lack of gains in adopting a commitment policy indicate that implementing a weighted inflation target is close to optimal.

\section{Conclusion}
In a standard two-sector New Keynesian model where policy makers maximise household welfare, the gains from commitment are negligibly small; or put differently, the stabilisation bias is close to zero. This result stems from the near-optimality of stabilising an appropriately weighted average of the two sectoral inflation rates. To the extent that a trade-off between stabilising the output gap and the inflation rate arises from sectoral asymmetries, attempts at devising policies to eliminate the inflation bias are unnecessary.

\newpage

\bibliographystyle{chicago}

\end{document}